%% file: main.tex
\documentclass[sigconf, nonacm, screen]{acmart}

\renewcommand\footnotetextcopyrightpermission[1]{} 
\AtBeginDocument{%
  \providecommand\BibTeX{{%
    \normalfont B\kern-0.5em{\scshape i\kern-0.25em b}\kern-0.8em\TeX}}}

\setcopyright{none}
\copyrightyear{2023}
\acmYear{2023}
\acmDOI{XXXXXXX.XXXXXXX}

\usepackage{natbib}
\usepackage{makecell}

\usepackage{graphicx}
\graphicspath{{./figures}}

\usepackage{listings}
\usepackage{xcolor}
\definecolor{darkgreen}{rgb}{0,0.6,0}
\usepackage{multirow} 
\usepackage{xspace}

\begin{document}

\title{Static Code Analysis in the AI Era: An In-depth Exploration of the Concept, Function, and Potential of Intelligent Code Analysis Agents}

\author{Gang Fan}
\email{fangang@antgroup.com}
\affiliation{%
\institution{Ant Group}
\city{Shenzhen}
\country{China}
}

\author{Xiaoheng Xie}
\email{xiexie@antgroup.com}
\affiliation{%
\institution{Ant Group}
\city{Shenzhen}
\country{China}
}

\author{Xunjin Zheng}
\email{zhengxunjin.zx@antgroup.com}
\affiliation{%
\institution{Ant Group}
\city{Shenzhen}
\country{China}
}

\author{Yinan Liang}
\email{lyn249877@antgroup.com}
\affiliation{%
\institution{Ant Group}
\city{Shenzhen}
\country{China}
}

\author{Peng Di}
\email{dipeng.dp@antgroup.com}
\affiliation{%
\institution{Ant Group}
\city{Hang Zhou}
\country{China}
}

\renewcommand{\shortauthors}{Trovato and Tobin, et al.}

\begin{abstract}
The escalating complexity of software systems and accelerating development cycles pose a significant challenge in managing code errors and implementing business logic. Traditional techniques, while cornerstone for software quality assurance, exhibit limitations in handling intricate business logic and extensive codebases. To address these challenges, we introduce the Intelligent Code Analysis Agent (ICAA), a novel concept combining AI models, engineering process designs, and traditional non-AI components. The ICAA employs the capabilities of large language models (LLMs) such as GPT-3 or GPT-4 to automatically detect and diagnose code errors and business logic inconsistencies. In our exploration of this concept, we observed a substantial improvement in bug detection accuracy, reducing the false-positive rate to 66\% from the baseline's 85\%, and a promising recall rate of 60.8\%. However, the token consumption cost associated with LLMs, particularly the average cost for analyzing each line of code, remains a significant consideration for widespread adoption. Despite this challenge, our findings suggest that the ICAA holds considerable potential to revolutionize software quality assurance, significantly enhancing the efficiency and accuracy of bug detection in the software development process. We hope this pioneering work will inspire further research and innovation in this field, focusing on refining the ICAA concept and exploring ways to mitigate the associated costs.
\end{abstract}



\received{28 September 2023}

\maketitle

\section{Introduction}
\input{Introduction.tex}

\section{Background and Related Work}

\input{Background.tex}

\section{Limitations and Motivations}
\input{Limitations}

\section{The Intelligent Code Analysis Agent}
\input{Definition}

\section{Agent Examples}
\input{Methodology}

\section{Evaluation and Field Test}
\input{Experiment}
\label{sec:Empirical Evaluation}

\section{Discussion}
\input{Discussion}

\section{Conclusion}
\input{Conclusion}


\clearpage
\bibliographystyle{plainnat}
\bibliography{references}

\end{document}

%% file: Introduction.tex
The rising complexity of software systems in our digitized world poses challenges in software engineering, particularly in managing code errors and implementing functional logic. Failure to address these issues promptly can result in system malfunctions, security vulnerabilities, and a negative user experience. The consequences include economic losses, reputation damage, and potential societal disruption. Timely identification and resolution of code errors and functional logic issues are critical for ensuring the reliability and proper functioning of software systems.

The traditional techniques of code error detection and functional logic verification, including code reviews, unit testing, and integrated testing, have been the cornerstone of software quality assurance for decades. However, they are inherently labor-intensive and often fall short when confronted with complex functional logic. These methods are heavily reliant on the expertise of the developers and the comprehensiveness of the test cases, both of which may not sufficiently account for all possible usage scenarios or intricate functional logic paths.

The rapid advancement of software development methodologies, such as Agile and DevOps, emphasizes the need for swift development and deployment cycles. This acceleration amplifies the necessity for an automated, intelligent, and efficient mechanism for the identification and rectification of code errors and functional logic inconsistencies. The urgency for such a system is particularly pronounced in the context of modern software systems, characterized by extensive codebases, multifaceted architectures, and sophisticated functional logic.

The contemporary digital landscape heralds the advent of LLMs and intelligent agent techniques, with examples such as AutoGPT standing out. These burgeoning technologies carve new trajectories for dealing with enduring challenges in software engineering. By harnessing the power of artificial intelligence and machine learning, these state-of-the-art technologies offer automated, intelligent, and efficient solutions, thereby breathing new life into longstanding, unresolved problems.

The notion of an intelligent agent represents a system that perceives its environment and takes actions to maximize its chances of achieving its goals. Applied to software engineering, these agents can be programmed to automatically identify and diagnose code errors and functional logic issues, potentially revolutionizing software development processes.

In this research, we introduce the concept of an ICAA — a novel idea that, to the best of our knowledge, has not been previously explored in existing literature. This agent leverages the power of AI models, engineering process designs, and traditional non-AI components to significantly improve bug detection in software development processes.

Our objective is to investigate and explore whether the application of intelligent agents, in conjunction with cutting-edge technologies like LLMs, could serve as a powerful tool to mitigate these entrenched issues. We propose a solution that leverages an intelligent agent grounded in state-of-the-art LLMs such as GPT-3 or GPT-4. This intelligent agent aims to automatically detect and diagnose code errors and functional logic inconsistencies through a sophisticated process of analysis and decision-making. We posit that our approach will considerably augment the field of software quality assurance, rendering it more efficient, precise, and cost-effective.

The concept of the ICAA is a broad and versatile one, with the potential to encompass a wide variety of designs and methodologies. In this work, we present one possible realization of this concept. Our implementation does not claim to be the only or definitive approach to designing a ICAA. Rather, it aims to provide an initial exploration of the concept's potential, demonstrating the power of combining AI models, engineering process designs, and traditional non-AI components.

Our primary contributions are as follows:

\begin{enumerate}
    \item \textbf{We propose the concept of the ICAA} This is, to our knowledge, the first work that introduces and explores this concept, marking a significant advancement in the field of code analysis.
    \item \textbf{We present a specific exploration of this concept}, demonstrating its potential to significantly improve bug detection. Our approach combines AI models, engineering process designs, and traditional non-AI components, effectively leveraging the strengths of each.
    \item \textbf{We provide empirical evidence}, not to show superiority, but to shed light on the potential of our approach. We present data and results that help illustrate the capabilities and possibilities of this new method in terms of bug detection accuracy and false-positive rate.
\end{enumerate}

By presenting this new concept and a successful exploration of it, we hope to inspire further innovations and varied approaches in the field of code analysis.

The remainder of this paper delves into the challenges in code error detection and functional logic verification, providing a comprehensive background and reviewing related work in this field. We then introduce the concept of the ICAA and showcase two specific implementations as examples. Following this, we present the results of our experiments, focusing on the performance of the intelligent agents in detecting code errors and functional logic issues. We conclude by discussing the implications of our findings, comparing this new method with traditional techniques, and suggesting potential directions for future research.

%% file: Background.tex
\subsection{Evolution of Software Development Methodologies}

The landscape of software engineering has undergone significant transformation over the past decades. Traditional waterfall development models have given way to Agile and DevOps, which promote rapid iterations, continuous integration, and seamless collaboration between development and operations teams \cite{beck2001agile, kim2016devops}. These methodologies have revolutionized software production, enabling the swift creation of robust and responsive software. However, they also introduce new challenges, particularly in maintaining code quality and ensuring the accurate implementation of business logic in large-scale, complex software systems.

\subsection{Automated Code Error Detection}

Automated error detection is a well-established area of research, with the objective to reduce the burden of manual code reviews and testing \cite{bessey2010few}. Several tools have been developed for this purpose, with static analysis tools, such as FindBugs for Java, being widely used \cite{ayewah2008using}. These tools inspect code without executing it, detecting potential bugs and vulnerabilities. Despite their benefits, such tools often produce false positives and struggle with the comprehension of complex business logic \cite{johnson2013why}, which adds noise and uncertainty to the development process. 
Dynamic analysis tools, which analyze code during or after its execution, offer an alternative approach. While these tools can detect errors that static analysis might miss, they also have limitations, such as performance overhead and difficulty in identifying non-deterministic bugs \cite{cornelissen2009systematic}.

\subsection{Functional Logic Verification}

Ensuring the accurate execution of intended functional logic is another vital aspect of software quality assurance. Traditional methods often involve manual inspection and extensive documentation, which can be both laborious and prone to human error \cite{balzarotti2010business}. Automated solutions, such as symbolic execution and model checking, have been proposed; however, these solutions often face challenges when dealing with large-scale, intricate systems due to scalability and state explosion problems \cite{clarke1999model, king1976symbolic}.

\subsection{LLMs and Intelligent Agents}

The advent of LLMs such as GPT-3 and GPT-4 has revolutionized automated software engineering, offering a profound understanding of language, context, and syntax \cite{brown2020language, openai2021gpt4}. These models, built upon extensive text training, have found diverse applications in coding tasks like code completion, code translation, and bug detection, where they have shown promising results \cite{feng2020codebert, karampatsis2020big, svyatkovskiy2020pythia}.

A significant stride in the LLM field was the introduction of CodeBERT by Microsoft and the more recent CodeLlama \cite{codellama}. These transformer-based models, trained on a vast corpus of code from GitHub repositories, have shown remarkable capabilities in tasks such as code summarization, code translation, and generating code from natural language prompts. Their performance underscores the potential of LLMs in understanding and generating code, opening up exciting opportunities for their incorporation into intelligent agents for static code analysis.

At the same time, the concept of intelligent agents — systems capable of interpreting their environment and taking actions to achieve their goals — has been garnering interest in software engineering. A case in point is AutoGPT\cite{githubGitHubSignificantGravitasAutoGPT}, an intelligent agent used in automated software testing, requirements analysis, and bug detection \cite{chen2018best, radford2019language}.

LangChain\cite{reactlm} is a powerful framework for developing applications powered by language models, serving as an intelligent agent in various code-related scenarios, including code analysis. LangChain enables developers to build context-aware applications that leverage the capabilities of language models to reason, provide code insights, and take appropriate actions based on the provided context.

Semantic Kernel\cite{semantic-kernel} is an open-source SDK that enables easy integration of AI services, such as OpenAI, Azure OpenAI, and Hugging Face, with programming languages like C\# and Python. By utilizing Semantic Kernel, developers can create AI applications that combine the strengths of both conventional programming and AI services.


The BabyAGI\cite{babyagi} project is an open-source implementation of an intelligent agent hosted on GitHub. This agent utilizes artificial intelligence (AI) capabilities, particularly natural language processing (NLP), to manage tasks. It makes use of the OpenAI API and vector databases like Chroma or Weaviate.

Intelligent agents\cite{talebirad2023multiagent} are software systems designed to perceive their environment, reason about it, and take appropriate actions to achieve specific goals. They possess advanced capabilities such as data analysis, decision-making, learning from experience, and interaction with users or other systems. 

\vspace*{8pt}

In light of these developments, this paper extends these lines of research by developing an intelligent agent that leverages an LLM for automated code error detection and functional logic verification. Our work aims to bridge the gap between the potential of LLMs and intelligent agents and their practical application in enhancing software quality assurance.

%% file: Limitations.tex
\subsection{Constraints of Static Code Analyzers}

Static code analysis tools serve as a pivotal element in ensuring software quality, facilitating the recognition of potential issues within code, including bugs, code smells, and security vulnerabilities. These instruments conduct their analysis on source code prior to execution, primarily concentrating on the code's syntactic and structural characteristics.

Nonetheless, these apparatuses have noteworthy limitations. Their reliance on rules often culminates in a high incidence of false positives and negatives\cite{bessey2010few}, thereby undermining trust in these tools and hampering productivity due to the necessity of manual review for flagged concerns. Additionally, traditional static analysis tools encounter difficulties in identifying bugs pertaining to functional correctness or business logic -- they are deficient in their capacity to apprehend the intricate meanings encapsulated in comments and the objectives of the developers, which is crucial to detect more elusive bugs.

\subsection{The Potential and Challenges of LLMs in Bug Detection}

The advent of LLMs, like GPT-3 \cite{brown2020language}, has instigated a significant shift in various AI fields. Their ability to comprehend and generate text akin to human communication has sparked the prospect of their application in code analysis and bug detection.

However, directly applying these models to code analysis and bug detection reveals several limitations. Though they have demonstrated abilities in tasks such as code generation, summarization, and even certain types of bug detection, their effectiveness in detecting bugs related to functional correctness and understanding code semantics is questionable\cite{johnson2013static}.

\subsubsection{Investigating the Performance of LLMs in Bug Detection}

To delve deeper into these limitations, an empirical investigation was conducted using GPT-3.5-turbo\cite{openai_blog_post}, a variant of GPT-3, to identify inconsistencies between code comments and their corresponding implementations. The dataset, composed of code blocks from various GitHub repositories, covered a range of programming languages and application domains.

The experiment entailed several stages:

\begin{enumerate}
\item A random sampling technique was employed on 540 million method code blocks in GitHub repositories with star ratings greater than 10, ensuring diverse programming language cases were considered. A total of 6000 data samples were selected, each representing the code content of a file.
\item The 6000 data samples were deduplicated, yielding 5712 unique data samples.
\item A second round of random sampling was conducted on these unique samples, drawing 500 samples per language, summing up to a total of 3000 data samples.
\item These samples were then split into code-comment pairs, selecting only those with non-empty method bodies and comments of more than 100 characters. This resulted in a total of 2380 data samples.
\item GPT-3.5-turbo was then tasked with analyzing these pairs for inconsistencies using a prompt template (Listing~\ref{lst:prompt1}).
\end{enumerate}

\begin{lstlisting}[language=Python, caption={Prompt template for analyzing code inconsistencies}, label={lst:prompt1}, basicstyle=\footnotesize]
Do not include any explanations in your responses. Find if there are inconsistencies between the source code and the comment.
Forget about the details and just focus on any inconsistency between the intention of the comment and the implementation of the code.
The generated result should be in JSON format, with the following fields:
"is_inconsistent": This should be true if the code and the comment are inconsistent, and false otherwise.
"explanations_and_suggestion": A description of why the code and the comment are inconsistent, and suggestions on how to fix it.
"fixed_comment": The revised comment, edited to remain consistent with the code.
"fixed_code": The revised code, edited to remain consistent with the comment.
The output format is as follows:
{{
    "is_inconsistent": <is_inconsistent>
    "explanations_and_suggestion": <explanations_and_suggestion>,
    "fixed_comment": <fixed_comment>
    "fixed_code": <fixed_code>
}}

Code:
```<Code Language>
<Code to Inspect>
```

Comment:
<Comment to Inspect>
\end{lstlisting}

\subsubsection{Analysis and Interpretation of Experimental Results}

The results revealed that while 53.4\%  of responses indicated inconsistencies, 46.1\% did not. A manual inspection and labeling of 440 randomly selected cases were conducted to understand these results better, as summarized in Table~\ref{tab:interpret_label}. The manual inspection revealed that a significant portion of the responses indicating no inconsistency were incorrect, despite a majority of responses indicating inconsistencies being accurate.

These findings underscore the potential of LLMs in detecting semantic inconsistencies—a task often challenging for traditional static code analyzers. An illustrative case of this potential is shown in Figure~\ref{lst:example_1}, where an inconsistency was detected in SaliencyMapper \cite{dabkowski2017real}, a Pytorch implementation of Real Time Image Saliency for Black Box Classifiers. The comment implies the need for an odd value for a particular parameter, but there is no corresponding enforcement in the code, unequivocally indicating a requirement that has been overlooked.

However, the results also highlight a considerable false-positive rate, surpassing what is typically observed with standard static analysis tools. This suggests that while LLMs are potent tools for understanding and generating text, their direct deployment for static code analysis remains a challenge.

\begin{lstlisting}[language=Python, caption={Prompt template for analyzing code inconsistencies}, label={lst:example_1}]
Code:
def gaussian_blur(_images, kernel_size=55, sigma=11):
    ''' Very fast, linear time gaussian blur, using separable convolution. Operates on batch of images [N, C, H, W].
    Returns blurred images of the same size. Kernel size must be odd.
    Increasing kernel size over 4*simga yields little improvement in quality. So kernel size = 4*sigma is a good choice.'''
    kernel_a, kernel_b = _gaussian_kernels(kernel_size=kernel_size, sigma=sigma, chans=_images.size(1))
    kernel_a = torch.Tensor(kernel_a)
    kernel_b = torch.Tensor(kernel_b)
    if _images.is_cuda:
        kernel_a = kernel_a.cuda()
        kernel_b = kernel_b.cuda()
    _rows = conv2d(_images, Variable(kernel_a, requires_grad=False), groups=_images.size(1), padding=(kernel_size / 2, 0))
    return conv2d(_rows, Variable(kernel_b, requires_grad=False), groups=_images.size(1), padding=(0, kernel_size / 2))

GPT-3.5-turbo Response:
The comment mentions that the kernel size must be odd, but the code does not enforce this. To make the code consistent with the comment, add a check to ensure that the kernel size is odd. Also, there is a typo in the comment (simga instead of sigma).

\end{lstlisting}

\begin{table}
    \centering
    \begin{tabular}{|c|c|c|} \hline 
         \textbf{GPT-3.5-turbo Response} &  \textbf{Manual Interpretation} & \textbf{Count} \\ \hline 
         \multirow{2}{*}{\textbf{Yes, they are consistent}} &  False Negative & 2 \\ \cline{2-3}
         &  True Positive & 218 \\ \hline 
         \multirow{1}{*}{\textbf{Unknown}} &  Ambiguous & 34 \\ \hline  
         \multirow{3}{*}{\textbf{No, they are not consistent}} &  False Positive & 147 \\ \cline{2-3}
         & True Negative & 31 \\ \cline{2-3}
         &  Indeterminate & 8 \\ \hline
    \end{tabular}
    \caption{Interpretation of GPT-3.5-turbo Responses}
    \label{tab:interpret_label}
\end{table}

The observed elevated false-positive rate in using LLMs for bug detection could potentially stem from a combination of several factors. Here are the three primary factors that might be contributing to the high false-positive rate:

\begin{enumerate}
\item \textbf{Syntax vs Semantics}: LLMs like GPT-3.5-turbo, while proficient in understanding syntax and grammar, may struggle with the semantics of code. Understanding code semantics requires a grasp of the underlying logic, various code structures, and developers' intentions, which might not be fully captured by these models.

\item \textbf{Contextual Gaps}: Code is often written within a specific context, which is crucial for understanding the rationale behind a comment or a code snippet. LLMs may not fully comprehend the broader context of the code, potentially leading to misinterpretations and false positives.

\item \textbf{Domain Knowledge and LLM Limitations}: Certain code comments require domain-specific knowledge, such as familiarity with specific algorithms, software frameworks, or application domains. A general-purpose language model like GPT-3.5-turbo might lack this domain-specific knowledge, leading to inaccuracies. Additionally, inherent LLM phenomena like hallucinations (generating information not present in the input) and randomness (the inherent unpredictability of the model) could also potentially contribute to false positives.
\end{enumerate}

Another issue observed was the uncontrollable output. We were unable to effectively control LLMs' output content, particularly the format, leading to frequent generation of malformed json responses. Furthermore, LLMs sometimes attempt to output the line number and the position of programmatic code. However, in most cases (147/186), these outputs were incorrect, posing challenges for potential integration and automatic processing with static code analyzers.

These observations underscore the need for a more refined approach to code analysis. While LLMs offer promising capabilities, their limitations, particularly in terms of false-positive rates and output control, pose significant challenges. There is a clear necessity for a solution that can leverage the strengths of LLMs while controlling their limitations effectively. This solution should also be able to adapt to automated processes and integrate smoothly with static code analyzers, addressing the challenges posed by current methodologies in bug detection.

To address these requirements, in the following section, we will introduce and define the concept of an ICAA and explore how it can enhance the accuracy and efficiency of code analysis tasks.

%% file: Definition.tex
The exploration of existing methodologies, namely traditional static code analysis tools and LLMs, has underscored distinct strengths and challenges. Static code analysis tools excel in analyzing the syntax and structure of the code, whereas LLMs demonstrate a promising ability to understand and generate human-like text. However, the former often struggles with semantic inconsistencies, and the latter exhibits high false-positive rates in bug detection.

In light of these insights, we propose the concept of an ICAA. This is not a specific design, but rather a strategic approach that aims to synthesize the strengths of both static code analysis tools and LLMs. It is an embodiment of an intelligent layer that organically integrates these tools, aiming to leverage their individual strengths while mitigating their limitations.

The ICAA is envisioned to harness the structural and syntactic analysis strengths of static code analysis tools. Simultaneously, it aims to capitalize on the semantic and contextual understanding provided by LLMs. The agent is thus designed to provide a more comprehensive understanding of the code, its semantics, and the intentions encapsulated in the comments.

Therefore, the motivation for proposing this concept is rooted in the desire to build upon the potentials and overcome the limitations of current methodologies. By employing an ICAA, we aim to provide a more robust, efficient, and holistic approach to code analysis.

\subsection{Distinguishing the ICAA from Traditional Static Code Analysis}

Traditional static code analyzers typically rely on fixed rules and patterns to identify potential issues within the source code, acting as a rule-based system that flags programming errors, bugs, or other issues according to predefined patterns. While these traditional tools have been instrumental in achieving a certain level of software quality, their effectiveness is inherently limited by the static set of rules they employ. This limitation potentially results in overlooking complex or rare bugs that fall outside these predefined patterns.

\subsection{ICAA: A Step Beyond Through Machine Learning Integration}

The ICAA transcends these limitations by ingeniously integrating machine learning algorithms into its operational framework. This integration allows the ICAA to continuously learn from the code it analyzes, facilitating the development of a sophisticated understanding of the makeup of "good" code and the patterns that might denote potential issues. 

\subsubsection{Learning from Inputs and Outputs}

The learning aspect of the ICAA is not limited to the code it analyzes. The ICAA's machine learning algorithms allow it to learn from both the inputs it receives and the outputs it generates during the analysis process. 

As the ICAA begins the analysis, it takes into account the input source code, its structure, and the various dependencies within the code. The machine learning algorithms allow the ICAA to learn from these inputs, incrementally building a model of the code's structure and behavior. This model evolves as the ICAA encounters new code patterns, enhancing its understanding of the overall code landscape.

Similarly, the ICAA learns from the outputs generated during the analysis process. Each time the ICAA flags a potential issue or evaluates the impact of a particular code pattern, it records and learns from the outcome. This continuous learning allows the ICAA to become more accurate and efficient over time in identifying potential issues and their associated impact.

\subsection{ICAA as a Decision-making Agent}

Beyond its analytical capabilities, the ICAA is an agent, signifying that it possesses decision-making capabilities. This represents a significant departure from traditional static code analysis tools, which are typically not designed to make decisions. The ICAA, however, can evaluate and prioritize potential issues based on their anticipated impact on the software's functionality. This capability allows developers to focus on resolving the most significant issues first, thereby improving the efficiency of the debugging process.

In summary, the ICAA is a dynamic, learning, and decision-making system within the realm of static code analysis. Its ability to continuously learn from the code it analyzes, and from the inputs and outputs during the analysis process, combined with its ability to make informed decisions about bug prioritization, represents a substantial advancement in the field of static code analysis. The introduction of the ICAA is anticipated to significantly enhance the efficiency of the debugging process, improve code quality, and contribute to the development of more robust software systems.

To better understand the practical implementation of an ICAA, we will introduce two example designs that highlight its capabilities and functionalities in next section.

%% file: Methodology.tex
Having introduced the concept of the ICAA, this section aims to provide concrete examples to enhance understanding of its design and functionality. We will present two distinct examples of ICAA, each showcasing a unique application: bug detection and intention-code consistency checking.

Our approach to designing these agents involves creating a specific chain of models or thought processes. This enables us to decompose the broader task of code analysis into multiple manageable components. Each agent's compositional structure encompasses both artificial intelligence and conventional components such as parsers. This innovative methodology illustrates how integrating machine learning can amplify the capabilities of traditional static code analysis techniques.

The first example we present is an agent designed for bug detection, highlighting how an ICAA can effectively identify and flag potential errors within a codebase. The second example, on the other hand, focuses on an agent tasked with ensuring intention-code consistency, demonstrating how an ICAA can assist in maintaining alignment between a developer's intentions and the actual implementation in code.

By detailing these examples, we aim to provide a clearer insight into the potential applications of an ICAA and how it can transform the landscape of code analysis. The following subsections provide an in-depth exploration of each example, elucidating our approach and the unique functionality of each agent.

\subsection{ReAct Bug Detection Agent}

\begin{figure*}[h]
  \centering
  \includegraphics[width=\textwidth]{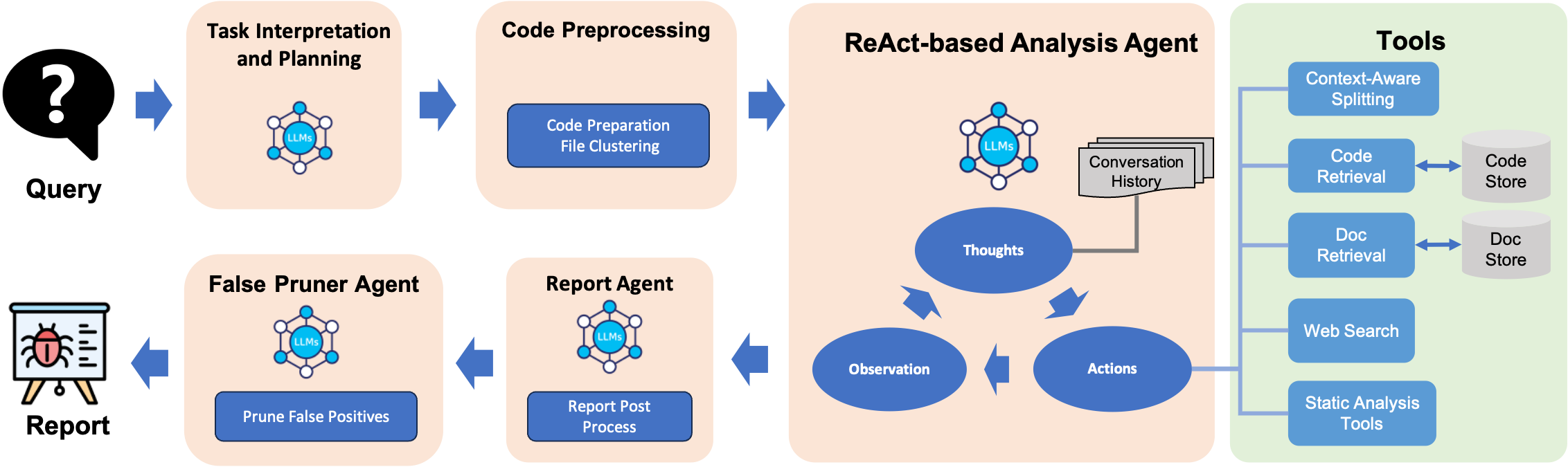}
  \caption{ReAct Bug Detection Agent}
  \label{fig:agent_version1}
\end{figure*}

We first outline the implementation details of an intelligent agent designed for code analysis, drawing inspiration from the ReAct framework \cite{reactlm}.
The agent forms a hierarchical structure composed of several sub-agents, each contributing to its overall capability. This is in accordance with the concept in AI that intelligent agents can be comprehended as a hierarchy of sub-agents working in unison.
This agent integrates a toolbox that includes context-aware splitting, code retrieval, documentation retrieval, web search, and static code analysis tools, all designed to aid developers in analyzing and comprehending code-related queries.
As depicted in Figure~\ref{fig:agent_version1}, the agent is composed of several components, the individual functionalities of which we will discuss in further detail.

\subsubsection{Task Interpretation and Planing}
In this step, the agent receives a user query as input. By using a prompt to query a LLM, the agent can interpret the task and devise a basic work plan for bug detection. This could involve, for example, examining a particular code snippet or defining the types of bugs of interest for a specific query. The returned work plan serves as a guideline for the entire analysis process.

\subsubsection{Code prepossessing}

At this stage, the agent downloads the appropriate code from the input and performs preprocessing. The primary task here involves categorizing the source files based on their types and programming languages. Additionally, we parse the source code and store it in a structured manner within a database. Thus, we store all the documents in one database and all code snippets in another. This arrangement facilitates efficient querying during the operation of the ReAct loop.

\subsubsection{ReAct-based Analysis Agent}
\label{sec:react agent}

The central part of the agent's design involves the intelligent agent that accepts an input query or a code snippet for analysis. This input could be a specific code-related question or a piece of code that needs thorough scrutiny.

The intelligent agent's functionality is deeply rooted in the ReAct framework, realized through its interactions with the Language Learning Model (LLM). This framework includes three critical components: Thoughts, Actions, and Observations.

\begin{enumerate}
    \item \textbf{Thoughts:}  The thoughts component serves as the brain of the agent, assessing the current situation and guiding the entire analysis process. It uses the conversation history - the comprehensive record of the dialogue between the agent and the LLM, including the sequence of prompts and responses - to inform its thinking. This context enables the agent to generate well-informed, contextually relevant thoughts.
    \item \textbf{Actions:} Actions are the steps taken by the agent based on its thoughts. These include sending prompts or code snippets to the LLM, processing the received responses, and utilizing tools such as static analyzers and parsers to interact more effectively with the environment. The primary goal of these actions is to enhance the agent's understanding of the code and improve its analysis capabilities.
    \item \textbf{Observations:} Observations are the feedback or responses received from the LLM. The agent uses these observations to refine its thoughts and decide on the next actions, ensuring a dynamic and responsive interaction process.
\end{enumerate}

The agent's operation significantly benefits from its interaction with the LLM, enabling it to generate insightful perspectives and boost its performance in code analysis tasks. By referencing previous dialogues, the agent avoids redundant queries, ensuring a more efficient and effective interaction process. This strategy not only conserves computational resources but also allows the agent to concentrate on the relevant aspects of the code analysis task.

\subsubsection{Actions and Tools}

Once the input is received, the agent employs its toolbox of tools to perform various actions based on the requirements. These actions include:

\begin{itemize}
    \item \textbf{Context-Aware Splitting Tool}: To address the challenge of analyzing large code snippets, we have incorporated a context-aware splitting tool. This tool considers the code context and intelligently splits large code snippets into smaller, manageable segments. By providing the agent with more granular input, this tool enhances the agent's performance in effectively analyzing and detecting bugs within large code snippets.
    
    \item \textbf{Code Retrieval Tool}: Implemented using a vector store, this tool converts the query into a vector and performs a search in the vector space. This method enables the agent to search and retrieve relevant code examples or snippets from code repositories like GitHub. To ensure accurate and meaningful vector representations, we carefully selected an advanced embedding model\cite{wang2022text}, which is trained on large-scale text corpora, capturing semantic relationships and contextual information. 

    \item \textbf{Documentation Retrieval Tool}: Similar to the Code Retrieval Tool, this tool uses vector space for efficient retrieval of relevant documentation or API references for various programming languages, frameworks, or libraries. It aids in providing information about usage, syntax, and available features related to the input query or code snippet.
        
    \item \textbf{Web Search Tool}: This tool performs searches to collect additional information from external resources related to the input query or code snippet. It can retrieve tutorials, articles, and forum discussions to supplement the analysis.
    
    \item \textbf{Static Code Analysis Tool}: This tool performs static code analysis on the provided code snippet. It checks for common coding errors, suggests improvements, provides code metrics, and identifies potential performance bottlenecks.
\end{itemize}

As the agent executes its toolbox actions, it engages in a dynamic thinking process that makes its approach adaptive and intelligent. This thinking process is iterative, where each cycle of actions and observations refines the agent's understanding of the input query or code snippet. As the agent interacts with the environment, it continuously updates its internal model, adjusts its actions, and incorporates new observations. This process allows the agent to evolve its strategies on the fly, making it capable of handling a wide array of code-related tasks with improved proficiency over time.

The agent's design is not only adaptive and intelligent but also anticipates potential enhancements. One such key enhancement is the integration of a Human-in-the-Loop (HITL) strategy\cite{human_in_the_loop}, where a human collaborator could be involved in the decision-making process, providing real-time inputs to the agent. The potential of this HITL strategy, although not currently implemented, could be a significant step towards creating more robust and efficient systems.


\subsubsection{Report Agent}
\label{sec:report}

The Report Agent is responsible for extracting information from the LLM responses and converting them into structured bug reports. While it is not inherently necessary to utilize LLMs in this step, their use in this design significantly enhances the conversion process. The LLM is particularly effective at dealing with the natural language outputs produced by the Consistency Checking Agent, converting these outputs into actionable bug reports. This process leverages sophisticated text extraction techniques and employs strategies such as 'guardrails', aimed at controlling the output schema of the LLMs. By defining certain parameters and constraints, the 'guardrails' technique ensures the LLM produces outputs that align with the expected format of the bug report, thereby facilitating a seamless translation process.




\subsubsection{False Pruner Agent}

The False Pruner Agent is an essential component in the system that meticulously inspects and refines the bug report produced by the Report Agent, with the explicit goal of reducing false positive results.

This agent can be implemented using either LLMs or static analysis tools. When utilizing LLMs, the agent applies their understanding of code semantics and patterns to identify potential false positives. On the other hand, when using static analysis tools, the agent cross-validates the initial results to recognize and remove false positives.

The decision between LLMs and static analysis tools depends on factors such as the complexity of the codebase, the availability of labeled datasets, and the desired accuracy level. Regardless of the chosen approach, the False Pruner Agent is designed to increase the reliability of the response.


\subsection{Code-Intention Consistency Checking Agent}
\label{sec:consistent agent}
\begin{figure*}[h]
  \centering
  \includegraphics[width=\textwidth]{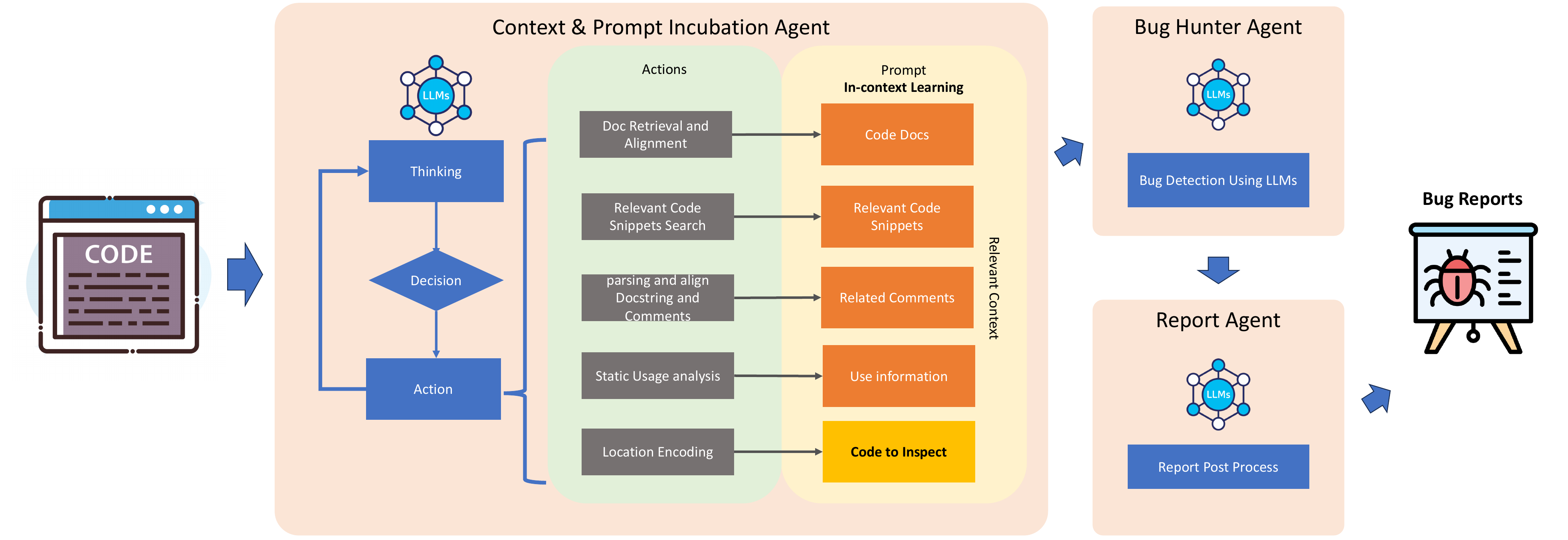}
  \caption{Code-Intention Consistency Checking Agent}
  \label{fig:agent_version2}
\end{figure*}

The second example design of the ICAA, as illustrated in Figure~\ref{fig:agent_version2}, targets the identification of inconsistencies between the intended functionalities of a code segment and its actual implementation. This design signifies an advancement over traditional static code analyzers by encompassing an understanding of the intent behind the code uses, semantics of comments, implications of documentation, and the naming conventions of functions and variables. This understanding is then juxtaposed against the actual code implementation to reveal potential discrepancies.

The motivation for this particular application of ICAA stems from the limitations of traditional static code analyzers. While these analyzers have been largely successful in detecting syntactical and certain types of logical errors, they often falter when it comes to inferring the intent behind a piece of code. Inferring intent involves understanding the semantics of not only the code but also the associated comments, documentation, and naming conventions of functions and variables. Traditional static code analyzers, being rule-based systems, struggle with such tasks due to their inherent limitations in dealing with semantic information. 

In contrast, ICAA, with its ability to utilize LLMs, can effectively decipher the subtle semantics embedded in the code and its associated artefacts. This enables the ICAA to infer the programmer’s intent and compare it against the actual implementation, thereby revealing any inconsistencies. This design choice makes ICAA particularly suitable for applications where understanding the semantics and intentions behind the code is crucial.

\subsubsection{Components}
This ICAA structure is composed of three primary components:

\begin{enumerate}
\item \textbf{Context \& Prompt Incubation Agent:} This component accepts code as input and dynamically navigates through a series of thinking, decision-making, and action steps to generate a context-sensitive prompt. These actions include the extraction and selection of relevant information from the code, an essential process given the window size limitations of current LLMs. Not all pertinent information is readily available or within the LLM's window size, necessitating an in-context learning approach to carefully select and prioritize the information that will best aid in bug detection. In-context learning, as defined by \cite{dong2023survey}, allows the model to adapt its predictions based on the specific sequence of tokens it's currently processing. This intricate, dynamic process is designed to cultivate a comprehensive understanding of the code's context, thereby enabling a more accurate and effective analysis.
  
    \item \textbf{Consistency Checking Agent:} The Consistency Checking Agent leverages prompt engineering techniques to query a LLM with the objective of detecting inconsistencies between the code's implementation and its context. This agent accepts input from the Context \& Prompt Incubation Agent, harnessing the advanced capabilities of the LLM to identify and analyze potential discrepancies. To facilitate the subsequent step, the LLM is configured to generate formatted outputs that can be efficiently parsed by the Report Agent. This strategic usage of formatted outputs simplifies the extraction and conversion process of the LLM’s responses, thereby enhancing the overall bug detection and reporting process.
  
 \item \textbf{Report Agent:} This component has been introduced in previous section \ref{sec:report}.
\end{enumerate}

\subsubsection{Location Encoding}

The "Location Encoding" step is an integral part of the ICAA process. This step involves inserting anchor symbols into each line of the source code, creating a form of 'geographical' landmarks within the code. The LLM can then refer to these landmarks to map its responses to the exact lines of code they relate to.

This approach addresses a common challenge when using LLMs for code analysis—maintaining accurate line numbers. Due to the complexity of source codes, LLMs often have difficulty mapping their outputs to the correct code lines. By embedding location information within the code, this complexity is managed more effectively.

Although this approach does consume additional tokens, it enhances the precision of locating potential issues in the code, thereby improving the overall effectiveness of the ICAA. As a result, the generated bug reports are more accurate, providing a more reliable basis for further analysis. Moreover, location encoding also improves the interoperability with other technical components. By providing a structured, location-based reference system within the code, other tools and systems can more easily interact with and understand the output from the LLM.

\subsubsection{Context \& Prompt Incubation Actions}
The Context \& Prompt Incubation Agent employs a multi-step procedure incorporating various actions. These include "Document Retrieval and Alignment", "Relevant Code Snippets Search", "Parsing and Aligning Docstrings and Comments", and "Static Usage Analysis". Each step contributes to constructing a comprehensive context for the LLM, enabling it to gain a deeper understanding of the code under analysis.

Contrary to a rigid procedure, the deployment of these actions is dynamically managed by a thinking-decision-action loop. This means there is no predefined order in which these steps occur. The agent, acting as the 'brain', determines whether or not to execute an action, the number of times an action is executed, and in what order. This dynamic decision-making process enhances adaptability and efficiency when dealing with varying code scenarios and complexities.

The process can commence with "Document Retrieval and Alignment", where relevant documents are identified and aligned for a coherent understanding of the code's context. "Relevant Code Snippets Search" can then locate key sections of the code, providing critical insights into the code's functionality. "Parsing and Aligning Docstrings and Comments" interprets and aligns docstrings and comments with the corresponding code sections, offering a narrative for the code's operation.
The "Static Usage Analysis" action leverages static analysis tools to decipher the code's structure and behavior. This process constructs a call graph and uncovers relevant code snippets, potentially revealing the programmer's intent. It provides deeper context to the LLM by analyzing the graph, function usage, and code similarity.

In-context learning is employed to effectively convey this context-rich information to the LLM. This approach enables the LLM to adapt its predictions based on the specific sequence of tokens it's currently processing, facilitating a more accurate and meaningful code analysis.

In the next section, we evaluate two example designs and compare them with the baselines to demonstrate the effectiveness and usefulness of the ICAA.

%% file: Experiment.tex
This section conducts an evaluation of the ICAA implementations. Rather than presenting an exhaustive assessment, our objective is to highlight the potential of the ICAA concept and to provide insights into the performance considerations of the two example implementations. In essence, our evaluations serve as field tests that illustrate the practical utility of the ICAA in real-world scenarios.
Our field tests are designed to address critical aspects of system performance, such as recall rate, false positive rate, and the cost associated with using intelligent agent-based approaches for identifying code errors.

Our evaluation strategy incorporates multiple benchmarks to provide diverse perspectives on our system's capabilities:
First, a comparative analysis with established baseline models offers a reference point for our system's performance. This comparison is not designed to assert superiority, but rather to clarify the relative strengths and limitations of our approach.
Next, an examination of actual source code repositories provides a realistic testbed. This allows us to evaluate the system's utility in detecting and addressing real-world bugs, providing insights into its practical performance.
Finally, a benchmarking exercise using code snippets containing different types of intentional errors serves to test our system’s bug detection capabilities in controlled conditions. This helps in appreciating the system's precision, recall, and token consumption metrics.

Through this evaluation, we aim to demonstrate the potential of the ICAA concept and provide insights into the performance considerations of our example implementations. By showcasing the practical application of our system in real-world scenarios, we contribute to the understanding and adoption of ICAA technology.

\subsection{Experiment Setup}

Our evaluation strategy uses baselines to illustrate the advantages of our ICAA approach. The primary baseline for comparison is the ChatGPT model, specifically the GPT-3.5-turbo variant, which is a significant part of our agent design and offers a valuable reference point.

In the baseline design, a prompt template is used to directly query the LLM for bug detection. The LLM, with training on an extensive corpus of code and natural language data, provides responses based on its learned capabilities.

Building upon this baseline, our ICAA approach optimizes the use of the LLM. Advanced prompt engineering techniques enable the ICAA to extract more relevant information and draw insights from the LLM's responses. This approach is both flexible and adaptable, permitting the ICAA to modify prompts dynamically and use the LLM's contextual understanding to make code recommendations, identify potential issues, and explain code behavior.

We opted for GPT-3.5-turbo over GPT-4 for our baseline due to its speed and cost-efficiency. However, we maintain that the comparison remains valid as long as the same base model is used. This comparison serves to emphasize the value of our agent design and its potential to enhance the system's capability and performance. Consequently, our evaluation strategy underscores the added value and enhancements brought about by the agent framework.

In the experiment we use to benchmark dataset:

\subsubsection{Benchmark Dataset - Non-Functional Bugs Dataset (\textbf{T1})}
\label{sec:benchmark}
The benchmark dataset, sourced from the "Non-Functional Bugs Dataset" available at \cite{ualberta-smr}, consists of real-world bugs related to non-functional requirements. It comprises 138 bugs from 67 open source projects, with 44 non-functional bugs in Python and 94 non-functional bugs in Java. The dataset provides scripts to process the data and includes separate folders for Java and Python projects, containing metadata, problem descriptions, and code improvement examples.

For our evaluation, we fed the benchmark dataset to our intelligent agent and measured its performance in detecting and resolving non-functional bugs. We analyzed the agent's false positive, recall, and token consumption in identifying these bugs, considering both language-specific and cross-language scenarios.

\subsubsection{Curated Test Suite (\textbf{T2})}
In addition to the benchmark dataset, we have meticulously curated a collection of 23 test cases extracted from closed issues on GitHub which focusing on API misuse scenarios. The API misuse test cases shed light on common mistakes and misuses of APIs encountered in software development. They provide examples of code errors resulting from incorrect API usage. 

To evaluate the performance of our intelligent agent on these test cases, we executed the agent on each individual test case and meticulously analyzed its ability to detect and address code errors. We quantified the false positive and recall of the agent in identifying code errors.

Through the evaluation of these two datasets, we were able to assess the strengths and limitations of our proposed approach in detecting and resolving both functional and non-functional bugs. The results obtained from these experiments yielded valuable insights into the effectiveness and applicability of our approach, particularly in terms of recall.
The curated test suite will be publicly available, enabling researchers and developers to access and utilize it for further evaluation and new research purposes.

All experiments are conducted on a laptop with a 6-core Intel(R) Core(TM) i7-9750H CPU @ 2.60GHz and 16GB of RAM.

\subsection{Evaluation of the False Positive Rate}
\vspace*{-8pt}
\begin{table}[h]
  \caption{Comparison of False-Positive Rates between ReAct Bug Detection Agent and Baseline}
  \centering
    \begin{tabular}{|c|c|c|}
      \hline
      Benchmark & ReAct Bug Detection Agent & Baseline \\
      \hline
      T1 & 66\% & 85\% \\
      \hline
  \end{tabular}
\end{table}
\vspace*{-8pt}
Our study demonstrates the improved performance of our ReAct Bug Detection Agent (see \ref{sec:react agent}) over a baseline. Both were tested on identical codebases, resulting in a direct comparison.

Our agents outperformed the baseline in bug detection accuracy, notably reducing the false-positive rate to 66\% from the baseline's 85\%. This highlights the enhancement in language model-based systems' capability achieved by our approach, despite a still relatively high false-positive rate.

Several factors contributed to the higher false-positive rate. These include the function check agent's incomplete caller recall, the use of an inadequately trained embedding model, and the choice of prompts and configuration settings. Potential improvements include enhancing caller recall, using more suitable models like CodeBERT, and fine-tuning prompts and settings.

These potential enhancements could decrease the false-positive rate and further boost our agents' performance, pointing towards a promising future for language model-based systems.
\subsection{Evaluation of the Recall}

\begin{table}[h]
  \caption{Recall Performance of the ReAct Bug Detection Agent on Benchmark 2}
  \centering
    \begin{tabular}{|c|c|}
      \hline
      Benchmark & ReAct Bug Detection Agent \\
      \hline
      T2 & 60.8\% \\
      \hline
  \end{tabular}
  \label{table:recall}
\end{table}

In order to evaluate the recall of the Code-Intention Consistency Checking Agent(see \ref{sec:consistent agent}), we have assembled a benchmark suite derived from real-world issues. This curated suite includes 23 test cases that primarily focus on API misuse. 

The ReAct Bug Detection Agent was executed on this dataset and the findings are presented in Table~\ref{table:recall}. Out of a total of 23 test cases, the agent was able to detect 14 issues, which equates to a recall rate of 60.8\%. This recall performance is not only comparable to that of the state-of-the-art techniques for detecting functional bugs\cite{sutingfunc}, but also underscores the potential and effectiveness of the proposed agent design in identifying bugs.

These results demonstrate that our Bug Detection Agent has the capability to accurately identify a significant proportion of real-world bugs. This, coupled with the fact that the agent's recall rate aligns closely with the performance of current state-of-the-art techniques, provides strong evidence of the efficacy and potential of the agent's design in the field of bug detection.

\subsection{Token consumption}

\begin{table*}[h]
  \caption{Comparison of Token Consumption Per Line for Baseline and Example Agents on Different Benchmarks}
  \centering
    \begin{tabular}{|c|c|c|c|}
      \hline
      Benchmark & Baseline & \makecell{ReAct Bug Detection Agent} & \makecell{Code-Intention Consistency \\  Checking Agent} \\
      \hline
      T1 &  600.3&  355.2&  417.8 \\
      \hline
      T2 &  510.0&  996.9&  834.6 \\ 
      \hline
  \end{tabular}
  \label{table:tokens}
\end{table*}

From the obtained data shown in Table~\ref{table:tokens}, the baseline implementation of the two agents exhibits similar token costs, ranging approximately from 355.2 to 996.9 tokens per line of code. This indicates a considerable cost. Given OpenAI's pricing at the time of writing this paper, which is \$0.0015 per 1K tokens for input and \$0.002 per 1K tokens for output, the average cost for analyzing each line of code is estimated to be between \$0.000812 and \$0.001032. Consequently, for a project comprising one million lines of code, the cost for a single analysis round is approximately \$1000. This expense could potentially hinder widespread adoption of these agents and may limit the potential use scenarios to those requiring only partial code analysis, such as Code Reviews. However, it is worth noting that the cost of language model usage has already been decreasing alongside the development of LLMs and the trend shows the price will continue to drop. Therefore, future research should focus on developing methods to further decrease token consumption to make these agents more cost-effective.

\subsection{Case Study}

\begin{lstlisting}[firstnumber=161, language=Java, numbers=left, caption={Error code}, label={lst:example1}]
@Override
public boolean onOptionsItemSelected(@NonNull MenuItem item) {
    if (!super.onOptionsItemSelected(item)) {
        switch (item.getItemId()) { 
            case R.id.clear_history_item:
                DBWriter.clearDownloadLog();
                return true;
            case R.id.refresh_item:
    ...
}
\end{lstlisting}

This is a case study that highlights a bug detected by our Intention-Code Consistency Agent. The bug was found in the test project of benchmark T2, specifically in the AntennaPod \footnote{\url{https://github.com/ByteHamster/AntennaPod}} repository hosted on GitHub.
Our agent reported that starting from line 165 (as shown in the above code snippet \ref{lst:example1}), where \textit{id.clear\_history\_item} is shown, the subsequent function call should invoke \textit{clearHistory} instead of \textit{clearDownloadLog}. This observation points to an inconsistency in the code implementation, suggesting that the intended behavior is to clear the history rather than the download log.
The reported inconsistency is indeed a case of incorrect implementation. The correct action should be to change \textit{R.id.clear\_history\_item} to \textit{R.id.clear\_log\_item}. From the perspective of detecting inconsistencies, this report is valid.

\begin{lstlisting}[language=java, numbers=left, firstnumber=154, caption={Relevant code}, label={lst:example2}]
    @Override
    public void onPrepareOptionsMenu(@NonNull Menu menu) {
        menu.findItem(R.id.episode_actions).setVisible(false);
        menu.findItem(R.id.clear_logs_item).setVisible(!downloadLog.isEmpty());
\end{lstlisting}

The reason behind our assumption that the LLM was able to identify this inconsistency lies in the fact that we utilized a static parser to provide the LLM with relevant code information. This included the filename \textit{DownloadLogFragment.java}, the class name \textit{DownloadLogFragment}, and the following relevant code snippet \ref{lst:example2}. The code snippet explicitly references another \textit{R.id} resource called \textit{R.id.clear\_log\_item}, which serves as a hint to the LLM regarding the existence of the correct \textit{R.id}.

In our investigation, we deliberately removed the contextual information from the prompt provided to the LLM. As a result, the LLM failed to detect the mentioned issues. We acknowledge that solely attributing the detection of such issues to the context fed to the LLM is not feasible, as LLMs pose significant challenges in terms of interpretability\cite{Linardatos2020ExplainableAA}.
Although the explainability of LLMs remains a challenge, we hypothesize that the absence of relevant contextual information significantly impacted the LLM's ability to identify the reported issues. We will continue our exploration in this direction to gain further insights and understanding.

\subsection{Summary and Conclusion}

Over the course of our evaluation, we have presented a comprehensive analysis of our ICAA implementations, focusing on their potential and performance in real-world scenarios. We conducted a series of field tests designed to assess crucial aspects of system performance such as recall rate, false positive rate, and the associated cost implications of using intelligent agent-based approaches for identifying code errors.


Our intelligent agent demonstrated improved performance over the baseline model in bug detection accuracy, reducing the false-positive rate to 66\% from the baseline's 85\%. Meanwhile, the recall rate stood at a promising 60.8\%, showcasing the agent's capability to accurately identify a significant proportion of real-world bugs. This aligns closely with the performance of current state-of-the-art techniques, providing strong evidence of the efficacy and potential of the agent's design in the field of bug detection.

However, despite these promising results, there are still challenges to overcome. The cost associated with token consumption is a significant consideration, and the average cost for analyzing each line of code could potentially hinder the widespread adoption of these agents. Future research should therefore focus on developing methods to further decrease token consumption to make these agents more cost-effective.


Building upon these findings, we now turn our attention to the discussion of the implications of the ICAA and explore potential avenues for further improvement and refinement of the ICAA in next section. 

%% file: Discussion.tex
In this section, we discuss the implications of our experimental results and their impacts on the field of software bug detection.

\vspace{4pt}
\textbf{Performance and Potential:} Our approach has demonstrated acceptable performance in detection rate and false-positive rate, suggesting that the design of our agent could be a viable direction for enhancing code error detection. During our evaluation, we did not compare our method with existing static analyzers, leaving the question of whether it can surpass these tools in terms of false positives and recall unresolved. This stands as a direction for future work. Despite this, our approach exhibits a tangible improvement over the baseline, further underlining its potential in the field of code error detection.

\vspace{4pt}
\textbf{Applicability:} Through our experiments, we have established that our approach, despite its challenges such as a high false-positive rate and substantial token consumption, holds promise. It presents several advantages over traditional static analysis, including the ability to handle incomplete code, easier integration, and a unique potential to detect complex bugs often missed by conventional methods. These case studies have not only enabled a deeper understanding of our system's operation in realistic scenarios but also underscored the potential of machine learning techniques in enhancing bug prediction based on past detections. While our approach is not fully practical in its current state, these insights suggest its potential applicability and room for future enhancements in real-world scenarios.

\vspace{4pt}
\textbf{Limitations:} Recognizing the potential limitations of our research is essential for understanding the validity of our findings. One potential threat to internal validity could arise from biases introduced by our experimental setup, such as the selection of specific codebases, tools, and our reliance on a specific LLM — GPT-3.5-turbo. Although GPT is currently state-of-the-art, the performance results may vary when different LLMs are used. Furthermore, our curated benchmark, while carefully selected, is not on a very large scale, which may limit the generalizability of our results. To bolster external validity, future work could involve testing our system on a broader range of codebases, using a variety of LLMs, and comparing its performance with a wider spectrum of bug detection tools.


An inherent feature of our AI-based system, which also represents a potential threat to reproducibility, is its non-deterministic nature. To address this, we designed experiments that include multiple runs of our AI agent on each codebase, averaging the results to derive a more robust performance estimate.

\vspace{4pt}
\textbf{Conclusion and Future work:}
In conclusion, our work serves as an initial step towards exploring new directions in functional correctness bug detection methodologies. We anticipate that our research will inspire further investigation and innovation, ultimately leading to the creation of more reliable and efficient software.
Our research opens several avenues for future work. Techniques to improve the deterministic behavior of our AI agent while retaining its ability to uncover bugs missed by traditional tools could be developed. Furthermore, extending our system to more programming languages would expand its usability and impact.

%% file: Conclusion.tex
In this research, we presented and discussed the concept of the ICAA. Our system signifies a step forward in the field of bug detection, showing the feasibility and potential of this approach. While there are still challenges to overcome and our system may not outperform all existing methods, the results obtained provide a clear indication of the possibilities in this novel direction of research.

Our exploration of the ICAA concept underscores its potential to bring about improvements in bug detection during software development processes. The effectiveness of our implementation suggests that this is just one instance of numerous possible designs within this broad concept, indicating that further exploration could lead to additional advancements in code analysis.

In conclusion, our work has highlighted a promising direction for enhancing static code analysis tools. We anticipate that our research will inspire further innovations and diversified approaches in the field of code analysis. This could ultimately lead to the development of more effective and efficient tools for software developers, thereby improving the software development process.